\documentclass[12pt,a4paper,dvips]{article}
\usepackage{ a4p}

\usepackage{amssymb,amsmath,graphicx}
\usepackage{cite,mcite}

\usepackage{ifthen}
\usepackage{l3_title}
\usepackage{Lep}
\usepackage{higgs_input}

\journalname{Phys. Lett. B}

\date{November 24, 2000}

\preprint{2000-145}
\Lep{2}
\begin{document}
\begin{titlepage}
  
  \title{Search for Neutral Higgs Bosons of the Minimal Supersymmetric
    Standard Model in $\boldsymbol{\epem}$ Interactions at
    $\boldsymbol{\sqrt{s}=192-202}$ $\boldsymbol{\GeV}$ }

  \author{The L3 Collaboration}
  
%
%
\begin{abstract}
  A search for the lightest neutral CP-even and the neutral CP-odd
  Higgs bosons of the Minimal Supersymmetric Standard Model is
  performed using 233.2\pb of integrated luminosity collected with the
  L3 detector at LEP at 
  centre-of-mass energies $192-202$\GeV. No signal is observed and   
  lower mass limits are given as a function of $\tanb$ for two scalar
  top mixing hypotheses. 
  For \tanb greater than 0.8, they are \mh $>$ 83.4\GeV and 
  \mA $>$ 83.8\GeV at 95\% confidence level.
\end{abstract}
  
%
%
\submitted
\vspace*{20mm}

\end{titlepage}
%
%
\section{Introduction}
\label{sec:intro}
 The Minimal Supersymmetric Standard
 Model (MSSM)~\cite{mssm_1} requires two Higgs
 doublets. This gives rise to
 five Higgs bosons: a charged scalar pair, two neutral CP-even, the
 lightest of which is called \h, and a neutral CP-odd, \A.
 The two most important production mechanisms in \epem collisions are: 
\begin{align}
  &\epemtohZ \label{eq:higgstrahlung} \\
  &\epemtohA. \label{eq:pairproduction}
\end{align}

 The cross section of the process~(\ref{eq:higgstrahlung}) is smaller 
 than that for the similar production of the Higgs
 boson in the Standard Model. This process is dominant at low values
 of \tanb ($\tan\beta\lesssim 5$), where \tanb is the ratio of the two
 Higgs vacuum expectation values. 
 The pair-production of Higgs bosons~(\ref{eq:pairproduction}) takes
 over at high values of \tanb.  

Previous searches for the h and A bosons were reported 
by L3~\cite{l3_1999_16} and other
experiments~\cite{opal_5}.
In this paper, we present the results of the search for the h and A bosons
using the data collected with the L3
detector~\cite{l3_1990_1}
at centre-of-mass energies $\rts=191.6-201.7\GeV$, corresponding to
233.2\pb of total integrated luminosity. 
The sensitivity to the production of MSSM neutral Higgs bosons 
is improved by combining the results of these analyses  
with our previous searches.

%
%
\section{Data and Monte Carlo Samples}
In 1999 the L3 detector collected data at LEP 
at average centre-of-mass energies $\sqrt{s}=191.6$\GeV, 195.5\GeV, 199.5\GeV
and 201.7\GeV, corresponding to the integrated luminosities
29.7\pb, 83.7\pb, 82.8\pb and 37.0\pb respectively.   

The cross sections of processes~(\ref{eq:higgstrahlung}) 
and~(\ref{eq:pairproduction}) 
 and the decay branching ratios of h and A are calculated using
 the HZHA generator~\cite{hzha}.
 For efficiency studies, Monte
Carlo samples of Higgs events are generated using
PYTHIA~\cite{pythia} and HZHA.  
2000 Monte Carlo events are generated for each mass hypothesis. For
\hA samples the masses, \mh and \mA, of the \h and \A bosons range
from 50 to 95\GeV in steps of 5\GeV. For \hZ samples \mh is chosen
in steps of 5\GeV from 50 to 95\GeV and in steps of 1\GeV from 95 to 110\GeV. 
For the background studies, the following Monte Carlo programs are used:
PYTHIA (\epemtoqqg, \epemtoZZ and
$\epem\!\rightarrow\!\Z\epem$), KORALW~\cite{koralw} 
(\epemtoWW), KORALZ~\cite{koralz} (\epemtotautau).
Hadron production in two-photon interactions is simulated with PYTHIA 
and PHOJET~\cite{phojet}.
EXCALIBUR~\cite{excalibur} is used for other four fermion final states. 
The number of simulated background
events for the most important background channels is more than 100
times the corresponding number of expected events.  

The L3 detector response is simulated using the GEANT 
program~\cite{geant}, which models the effects of energy
loss, multiple scattering and showering in the detector. The GHEISHA
program~\cite{gheisha} is used to simulate hadronic interactions in
 the detector. Time dependent inefficiencies are also taken
 into account.

%
\section{Event Selection}
For the hA production, the following decay modes are considered:
$\mathrm{hA\rightarrow \mathrm{b\bar{b}b\bar{b}}}$, 
$\mathrm{hA\rightarrow \mathrm{b\bar{b}\tau^+\tau^-}}$ and 
$\mathrm{hA\rightarrow \mathrm{\tau^{+}\tau^{-}b\bar{b}}}$.
In the case of hZ, four event topologies covering approximately
98\% of possible final states, are considered: $\mathrm{q\bar{q}q\bar{q}}$,
$\mathrm{q\bar{q}\nu\bar{\nu}}$, 
$\mathrm{q\bar{q}l^+l^-(l=e,\mu,\tau)}$ and
$\mathrm{\tau^+\tau^- q\bar{q}}$. The searches in channels with hadronic
decays of the \h boson are optimised for 
the dominant $\mathrm{h\rightarrow b\bar{b}}$ decay channel.
The analyses $\mathrm{q\bar{q}\nu\bar{\nu}}$ and 
$\mathrm{q\bar{q}l^+l^-(l=e,\mu)}$ are the same as devised for the
Standard Model Higgs search~\cite{l3_sm_higgs_00_paper}.
  
A common selection is applied to both the \hA and \hZ searches.
In the four-jet channel, it mainly
reduces the two-photon interaction background while keeping the signal 
efficiency high, then a neural network is used to build a
discriminating variable. 
In the tau channel, first a selection is devised, 
then an optimal variable based on a
likelihood approach is defined.    
%
%
\subsection{$\boldsymbol{\hAtobbbb}$\, and $\boldsymbol{\hZtobbqq}$\, Selection}
\label{sec:bbbb}
The signature of both the \hAtobbbb and \mbox{\hZtobbqq} final states
is four hadronic jets and the presence of b-hadrons.
The dominant backgrounds come from $\mathrm{q\bar{q}(\gamma)}$ 
production and hadronic decays of \W and \Z pairs.  

High multiplicity events are selected and their visible energy,
$\mathrm{E_{vis}}$, is required to be greater than $0.6\sqrt{s}$ and
less that $1.4\sqrt{s}$.
Events with perpendicular imbalance greater than
0.35$\mathrm{E_{vis}}$ or with a lepton whose energy exceeds 65\GeV  
are rejected to suppress semileptonic W pair decays.
Initial state radiation events are further suppressed by requiring 
{$P_{\mathrm{mis}}^{\mathrm{L}}/(m_{\mathrm{vis}}-\mZ)<0.4$}, where 
$P_{\mathrm{mis}}^{\mathrm{L}}$ 
is the longitudinal component of the missing momentum, 
$m_{\mathrm{vis}}$ is the 
visible mass and \mZ the \Z boson mass.
The remaining events are then forced into
four jets using the DURHAM algorithm~\cite{DURHAM} and a
kinematic fit requiring energy and momentum conservation (4C) is performed.
The number of expected and observed events in
the data, together with the signal efficiencies for the
selection cuts are listed in Table~\ref{tab:4jet_eff}.

\begin{figure}[htb]
  \begin{center}
    \includegraphics*[width=0.9\textwidth,bb=12 201 625 1018]{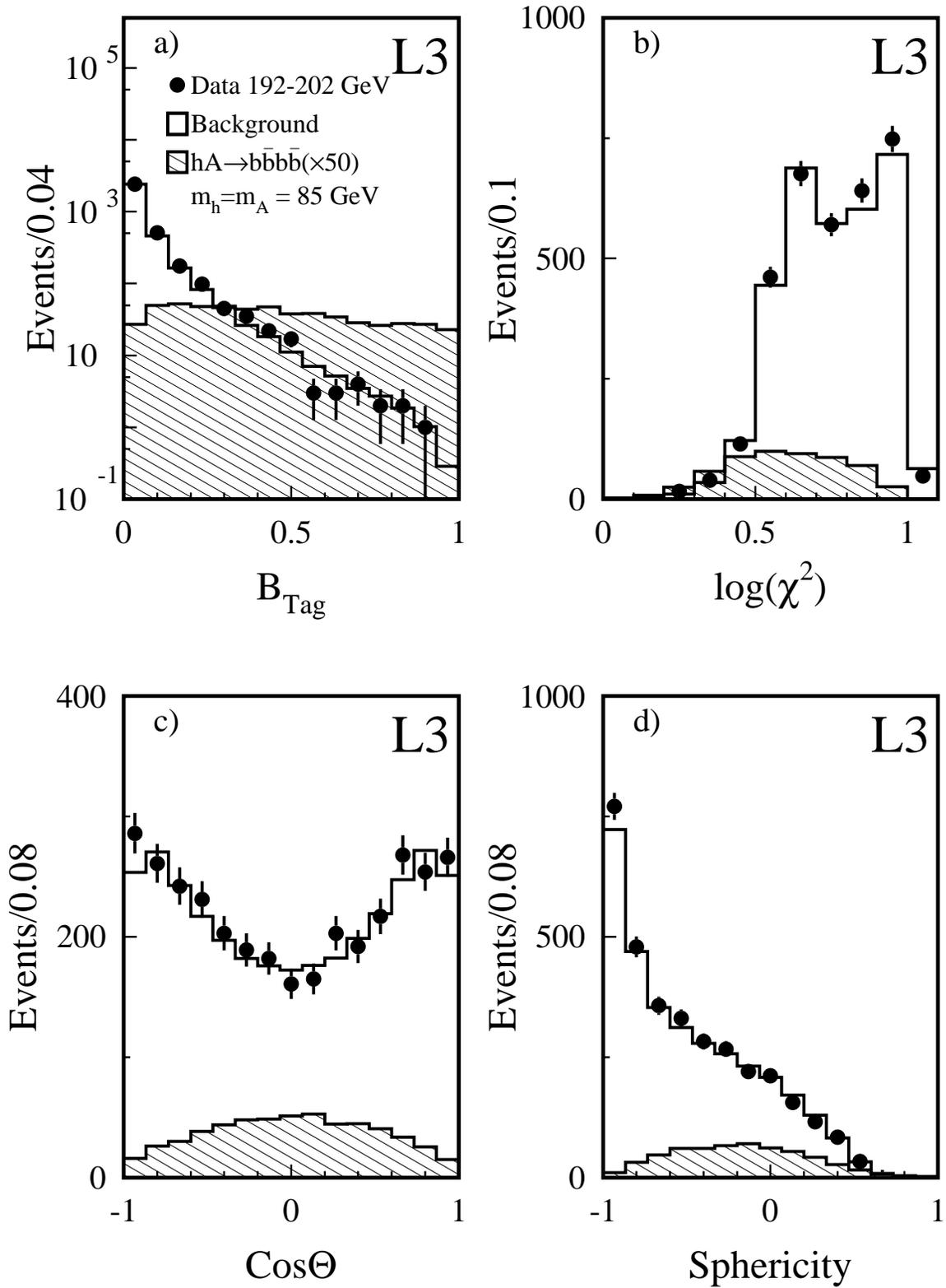}
    \caption{Distributions of a) $\rm{B_{Tag}}$, 
      b) logarithm of the $\chi^2$ variable,
      c) cosine of the polar angle $\Theta$ of the Higgs boson 
      and d) sphericity after the four-jet selection.
      The points represent 
      the data at $\rts=192-202\GeV$, the open histograms are
      the expected Standard Model backgrounds, and the hatched 
      histograms are the
      expected $\mathrm{hA \rightarrow b\bar{b}b\bar{b}}$ signal
      scaled by a factor 50, for $\mh=\mA=85\GeV$ at $\tanb=30$.} 
    \label{fig:var_4b}
  \end{center}
\end{figure}

After the selection, discriminating variables
are combined into a feed forward neural network~\cite{jetnet_1},
with one hidden layer and three output nodes.
The inputs include the probability that the jets contain 
b-quarks~\cite{l3_1997_18}, hereafter termed $\rm{B_{Tag}}$, event
shape variables and mass information.
The information on the event shape includes
 the event sphericity, the value of the DURHAM 
jet resolution parameter for which the event is resolved from three
to four jets,
the longitudinal component of the missing momentum and the event thrust.
The mass information is summarised in a $\mathrm{\chi^2}$ variable 
defined as:  
\begin{equation}
\chi^2(m_X,m_h)=
(\Sigma^{min}_{ij}-(m_X+m_{\mathrm h}))^2w_{\Sigma}+
(\Delta^{min}_{ij}-|m_X-m_{\mathrm h}|)^2w_{\Delta}~,
\end{equation}
where ${m_X}$ is either $m_{\mathrm Z}$ or $m_{\mathrm A}$. 
The pairing used gives the minimum value for
the difference squared $ (\Delta_{ij} -|m_X-m_{\mathrm h}|)^{2}$ between
the measured and expected dijet mass differences;
$\mathrm{\Sigma_{ij}^{min}}$ is the corresponding
sum of the dijet masses.
The weights, $w_{\Sigma}$ and $w_{\Delta}$, are derived from the
mass resolutions, and their ratio is 3/5.
In this way, the shape of the neural network output is made almost independent
of the mass hypothesis. 
The polar angle of the Higgs boson, $\mathrm{\Theta}$,
is also used as input for the \hA analysis.
It gives additional  separation between the hA signal and $\mathrm{W^+W^-}$
background,
due to the different spins of the W and the Higgs bosons.  
The distributions of the discriminating input variables
are shown in Figure~\ref{fig:var_4b}.

\begin{figure}[htb]
  \begin{center}
    \includegraphics*[width=0.9\textwidth,bb=8 200 628 1020]{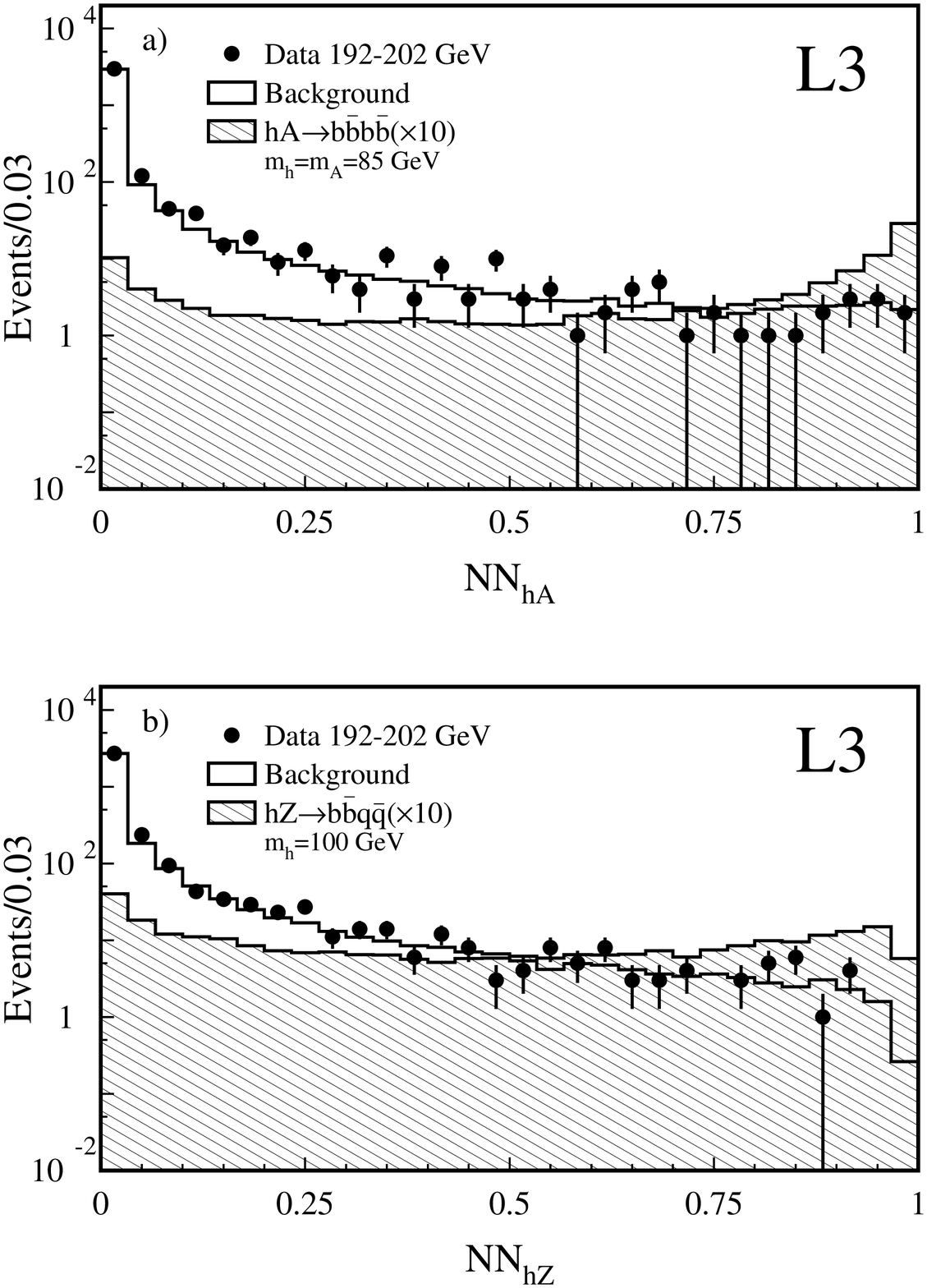}    
    \caption{Distributions of the discriminating variable in the four
    jet channel for 
      a) $\mathrm{hA\rightarrow{b\bar{b}b\bar{b}}}$ and 
      b) $\mathrm{hZ\rightarrow{b\bar{b}q\bar{q}}}$. 
      The points show the data collected at
      $\sqrt{s}=192-202\GeV$, 
      the open histograms are the expected Standard Model backgrounds and  
      the hatched histograms are the expected signals scaled by a
      factor 10. 
      The discriminating variables are constructed assuming
      equal mass hypothesis $\mh=\mA=85\GeV$ at $\tanb=30$
      in a) and $\mh=100\GeV$ at $\tanb=1$ in b).}
    \label{fig:fv_4b}
  \end{center}
\end{figure}

The Neural Network has three output variables which correspond 
to the signal,  ${O_{Higgs}}$,
the $\mathrm{q\bar{q}(\gamma)}$ final state, ${O_{qq}}$, and
the $\mathrm{W^+W^-}$ final state hypotheses, ${O_{WW}}$.
The discriminating variable is then obtained 
from the combination 
${NN} = {O_{Higgs}}\times(1 - {O_{qq}})\times(1 - {O_{WW}})$.
Two different neural networks are used for the
\hA and \hZ analyses, and the events are classified as \hA or \hZ 
according to the largest value of the discriminating variable.
The neural network for the \hZ is optimised for $m_{\mathrm h}=100\GeV$ 
at $\sqrt{s}=192-196\GeV$
and $m_{\mathrm h}=105\GeV$ at $\sqrt{s}=200-202\GeV$. 
For the \hA search, the neural network is optimised for 
$m_{\mathrm h}=m_{\mathrm A}=85\GeV$. 
Examples of the discriminating variable distributions of
the two analyses are shown in Figure~\ref{fig:fv_4b}. 
Good agreement is observed between the data and the 
expected background.  
An independent analysis based on a likelihood approach~\cite{l3_1998_16}
validates the present results.

\begin{figure}[htb]
  \begin{center}
    \includegraphics*[width=0.9\textwidth,bb=8 200 628 1020]{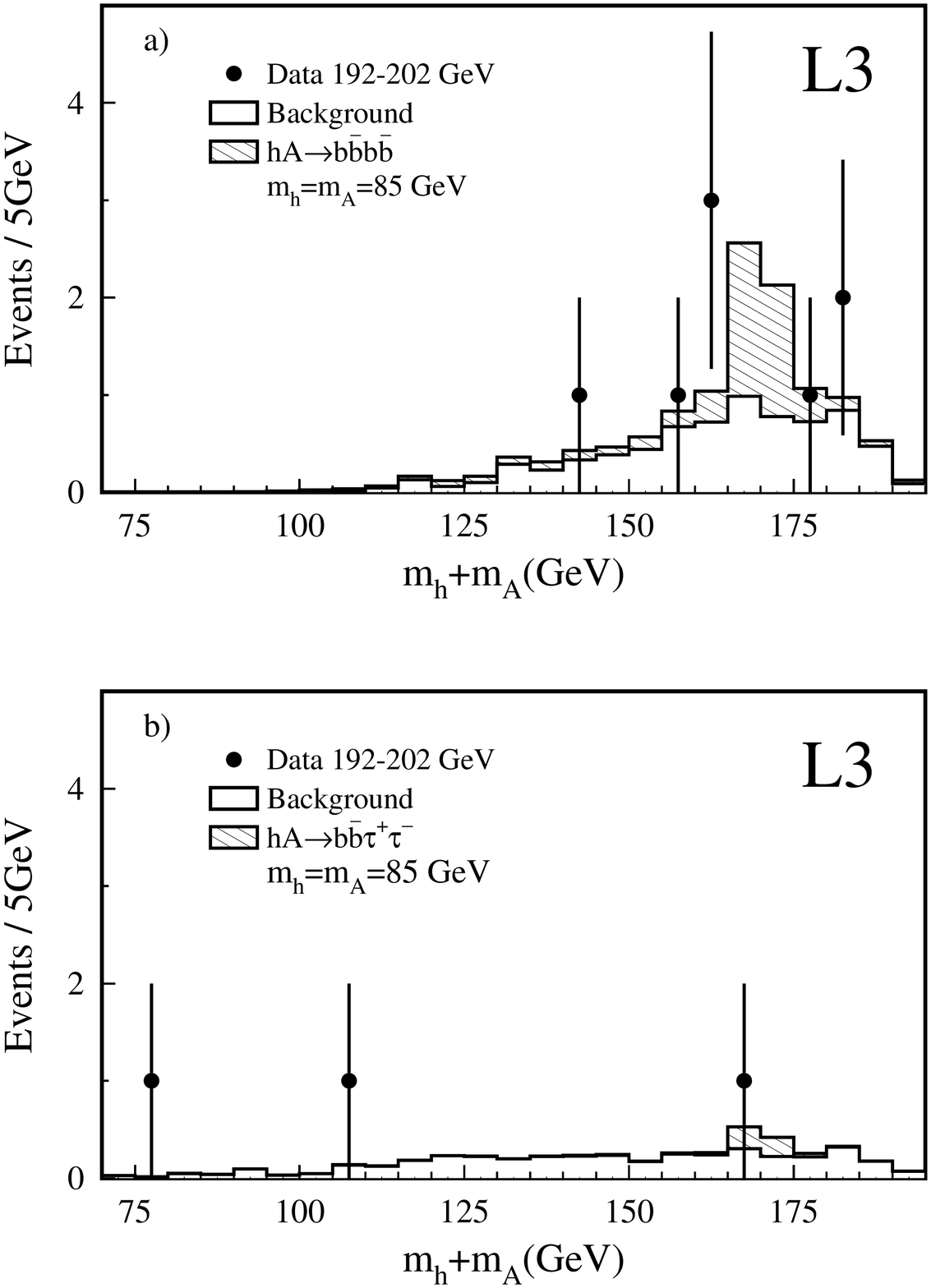}
    \caption{
             Distributions of a) the sum of the reconstructed dijet masses 
             in the $\hAtobbbb$ channel after
             the cut $NN>$0.9,
             b) the sum of the dijet and the ditau masses in the 
             $\hAtobbtt$ channel after a cut on the $\rm{B_{Tag}}$.
             The points represent the data collected at
             $\sqrt{s}=192-202\GeV$, the open histogram is the
             expected Standard Model background and the hatched
             histogram is the expected signal for $\mh=\mA=85\GeV$ at $\tanb=30$.}
    \label{fig:mass}
  \end{center}
\end{figure}

The highest signal
sensitivity region corresponds to large values of ${NN}$. 
For illustrative purposes, in Table~\ref{tab:4jet_eff} we report the
number of the observed and expected events selected after the cuts 
${NN}>$0.9 for the \hA analysis, and ${NN}>$0.5 for the \hZ
analysis.
The mass distributions for the \hA search after the cut
${NN}>$0.9 is shown in Figure~\ref{fig:mass}a.  

\begin{table}[htb]
  \begin{center}
    \begin{tabular}{|c|rrr|rrr|}
    \hline
 & & 192\GeV & & & 196\GeV & \\
\hline  
Cut & Selection & $NN_{\mathrm{hA}}$$>$0.9 & $NN_{\mathrm{hZ}}$$>$0.5 & 
Selection & $NN_{\mathrm{hA}}$$>$0.9 & $NN_{\mathrm{hZ}}$$>$0.5 \\
\hline
Data & 430 & 2 & 4 & 1186 & 3 & 33 \\
\hline
MC & 433.6 & 1.0 & 9.5 & 1197.5 & 3.2 & 26.0 \\
\hline
$\mathrm{q\bar{q}}$ & 201.2 & 0.4 & 3.0 & 531.3 & 1.2 & 7.1 \\
$\mathrm{W^+W^-}$ & 217.9 & 0.2 & 4.2 & 622.3 & 0.7 & 11.9 \\
$\mathrm{ZZ}$ & 14.5 & 0.4 & 2.3 &43.9 & 1.3 & 7.0  \\
\hline
$\mathrm{\varepsilon_{hA}}$ & 95.0\% & 42.0\% & 65.9\% & 95.0\% & 42.0\% &
 65.9\%  \\
\hline
$\mathrm{\varepsilon_{hZ}}$ & 93.7\% & 12.6\% & 52.7\% & 93.7\% & 12.6\% &
 52.7\% \\
\hline\hline
 & & 200\GeV & & & 202\GeV & \\  
\hline
Cut & Selection & $NN_{\mathrm{hA}}$$>$0.9 & $NN_{\mathrm{hZ}}$$>$0.5 & 
Selection & $NN_{\mathrm{hA}}$$>$0.9 & $NN_{\mathrm{hZ}}$$>$0.5 \\
\hline
Data & 1198 & 2 & 9 & 506 & 1 & 6\\
\hline
MC & 1118.4 & 2.0 & 9.1 & 507.8 & 1.0 & 4.8 \\
\hline
$\mathrm{q\bar{q}}$ & 478.3 & 0.7 & 2.5 & 215.0 & 0.3 & 1.5 \\
$\mathrm{W^+W^-}$ & 595.4 & 0.3 & 3.3 & 271.8 & 0.2 & 1.7 \\
$\mathrm{ZZ}$ & 44.7 & 1.0 & 3.3 & 21.0 & 0.5 & 1.6 \\
\hline
$\mathrm{\varepsilon_{hA}}$ & 94.1\% & 37.9\% & 37.9\% & 95.9\% & 38.3\% & 40.4\% \\
\hline
$\mathrm{\varepsilon_{hZ}}$ & 91.0\% & 5.7\% & 40.9\% & 92.3\% & 5.0\% & 36.3\% \\
\hline
    \end{tabular}
    \caption{Number of events observed and expected in the 
      four-jet channels, after the selection and after cuts
      on the discriminating variables, $NN_{\mathrm{hA}}$ and 
      $NN_{\mathrm{hZ}}$. The \hA signal
      efficiencies are quoted for $\mA=\mh=85\GeV$. The \hZ 
      signal efficiencies correspond to  
      $\mh=100\GeV$ at $\sqrt{\mathrm{s}}=192-196\;\GeV{}$ and
      $\mh=105\GeV$ at $\sqrt{\mathrm{s}}=200-202\;\GeV{}$.}
    \label{tab:4jet_eff}
  \end{center}
\end{table}

%
%
\subsection{$\boldsymbol{\hA}\!\boldsymbol{\rightarrow}\!\boldsymbol{\bbbar}\pmb{\tau}^{\boldsymbol{+}}\pmb{\tau}^{\boldsymbol{\, -}}$, $\boldsymbol{\hZ}\!\boldsymbol{\rightarrow}\!\boldsymbol{\bbbar}\pmb{\tau}^{\boldsymbol{+}}\pmb{\tau}^{\boldsymbol{\, -}}$\, and $\boldsymbol{\hZ}\!\boldsymbol{\rightarrow}\!\pmb{\tau}^{\boldsymbol{+}}\pmb{\tau}^{\boldsymbol{\, -}}\boldsymbol{\qqbar}$ Selections}
\label{sec:bbtt}
 The signatures of $\hAtobbtt$
 \footnote{Both the decay modes 
 (h$\rightarrow \mathrm{b}\mathrm{\bar b}$, A$\rightarrow\tau^+\tau^-$)
 and
 (h$\rightarrow\tau^+\tau^-$, A$\rightarrow \mathrm{b}\mathrm{\bar b}$) 
 are considered.}, 
 \hZtobbtt or \hZtottqq
events are a pair of taus accompanied by two hadronic jets.  
For each
of the channels \hA and \hZ  an analysis is optimised
based either on the tau identification or on
the event topology by requiring four jets 
with two of them being narrow and of low multiplicity.
The main
background results from W-pair decays containing taus.

The hZ analysis is similar to the one
described in detail in Reference~\citen{l3_sm_higgs_99_paperpub}.
The \hA selection is optimised for lower Higgs masses
by omitting the cuts on the opening angles of
the jets and tau pairs and on the invariant mass of
the tau pair, $m_{\mathrm\tau\tau}$.
The  invariant mass of the hadronic jets,
$m_{\mathrm q\mathrm q}$, must be between 5\GeV and
125\GeV.  
The ratio of the sum
of the energies of the tau decay products over the sum of the energies
of the jets is required
to be less than one and the value of the missing momentum vector in
the rest frame of the Higgs must be less than 40\GeV.
Finally, the cosine of the polar angle
of the Higgs boson, $|\cos\Theta|$, has to
be less than 0.8.  
The number of events observed, the
number expected from background processes, and the signal efficiency
for the two selections are listed in Table~\ref{tab:bbtt_eff}.

\begin{figure}[htb]
  \begin{center}
    \includegraphics*[width=0.9\textwidth,bb=11 199 622 1016]{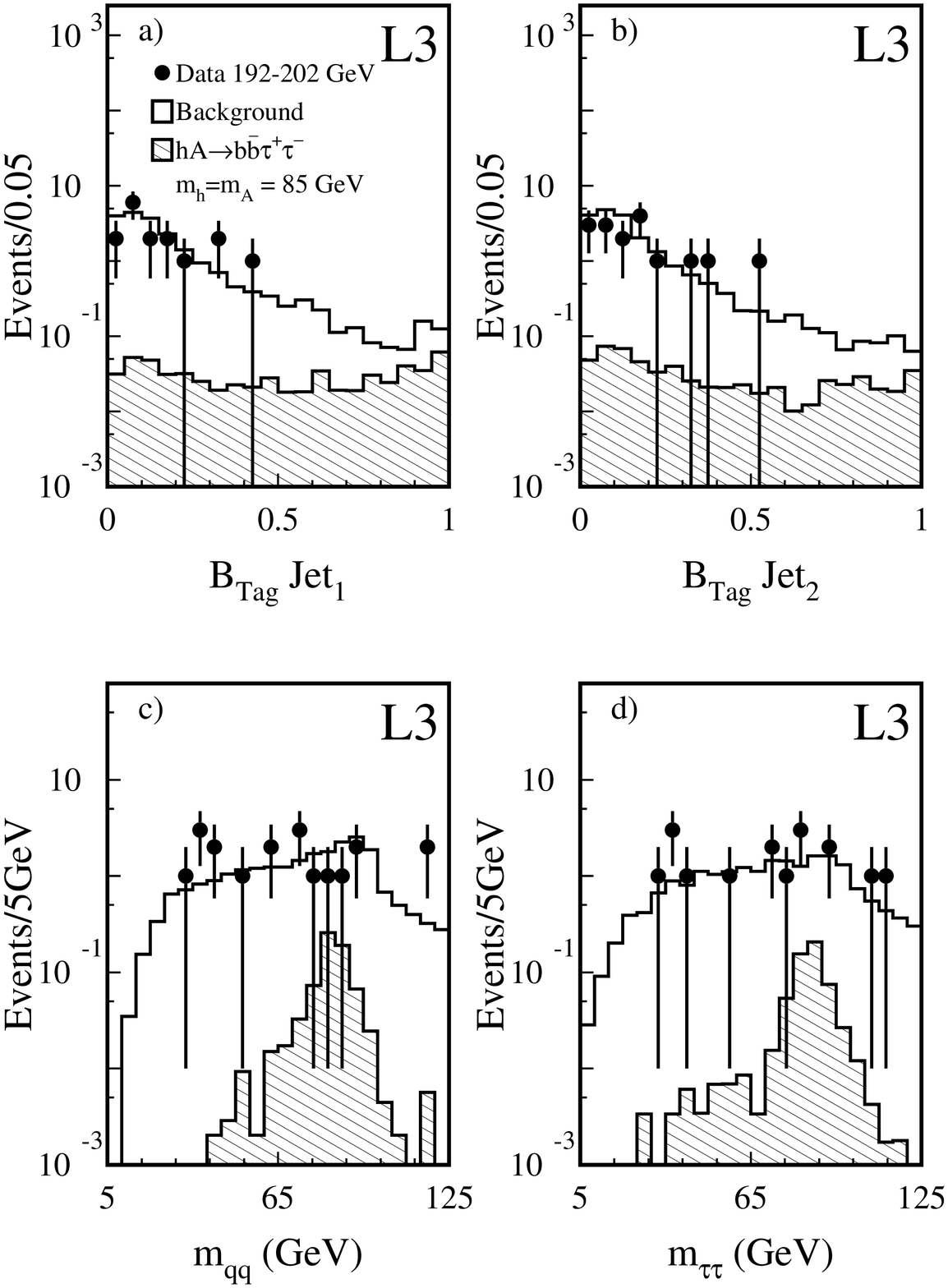} 
    \caption{The distributions for the $\hAtobbtt$
      search of a) the $\rm{B_{Tag}}$ for the most energetic hadronic
      jet and
      b) the least energetic hadronic jet, c) the reconstructed mass 
      for the hadronic
      system, and d) the reconstructed mass for the leptonic system.
      The points are the $\rts=192-202\GeV$ data, the open histograms the
      expected Standard Model background and the hatched histograms 
      the expected
      $\hAtobbtt$ signal for $\mh=\mA=85\GeV$ at $\tanb=30$.}
    \label{fig:last_vars_bbtt}
  \end{center}
\end{figure}

For each event class $j$ ($\mathrm{ZZ}$, $\mathrm{W^+W^-}$, 
$\mathrm{q\bar{q}}$, $\mathrm{Ze^+e^-}$, \hA,
\hZ), a probability function, $f_{j}^{i}$, is constructed,
where $i$ denotes the variables considered. These are the
$\rm{B_{Tag}}$ for each
hadronic jet, $m_{\mathrm q\mathrm q}$ and $m_{\mathrm\tau\tau}$. 
They are presented in Figure~\ref{fig:last_vars_bbtt}.
The probability, $p_{j}^{i}$, of an
event to belong to class $j$, based on the value of the variable $i$,
is then defined as  
\begin{equation}
p_{j}^{i}=
\frac{f_{j}^{i}}{\sum_{k}{f_{k}^{i}}}~.
\end{equation}
Finally, the probabilities for the individual variables are combined
by calculating the likelihood that the event belongs to the either signal
class:
\begin{equation}
F_{{\mathrm h}{\mathrm A}}=
\frac{\prod_{i}p_{{\mathrm h}{\mathrm A}}^{i}}{\sum_{k}{\prod_{i}p_{k}^{i}}}
~~~~~~{\rm and}~~~~~~
F_{{\mathrm h}{\mathrm Z}}=
\frac{\prod_{i}p_{{\mathrm h}{\mathrm Z}}^{i}}{\sum_{k}{\prod_{i}p_{k}^{i}}}~.
\end{equation} 

\begin{center}
\begin{figure}[htb]
  \begin{center}
    \includegraphics*[width=0.9\textwidth,bb=9 605 625 1019]{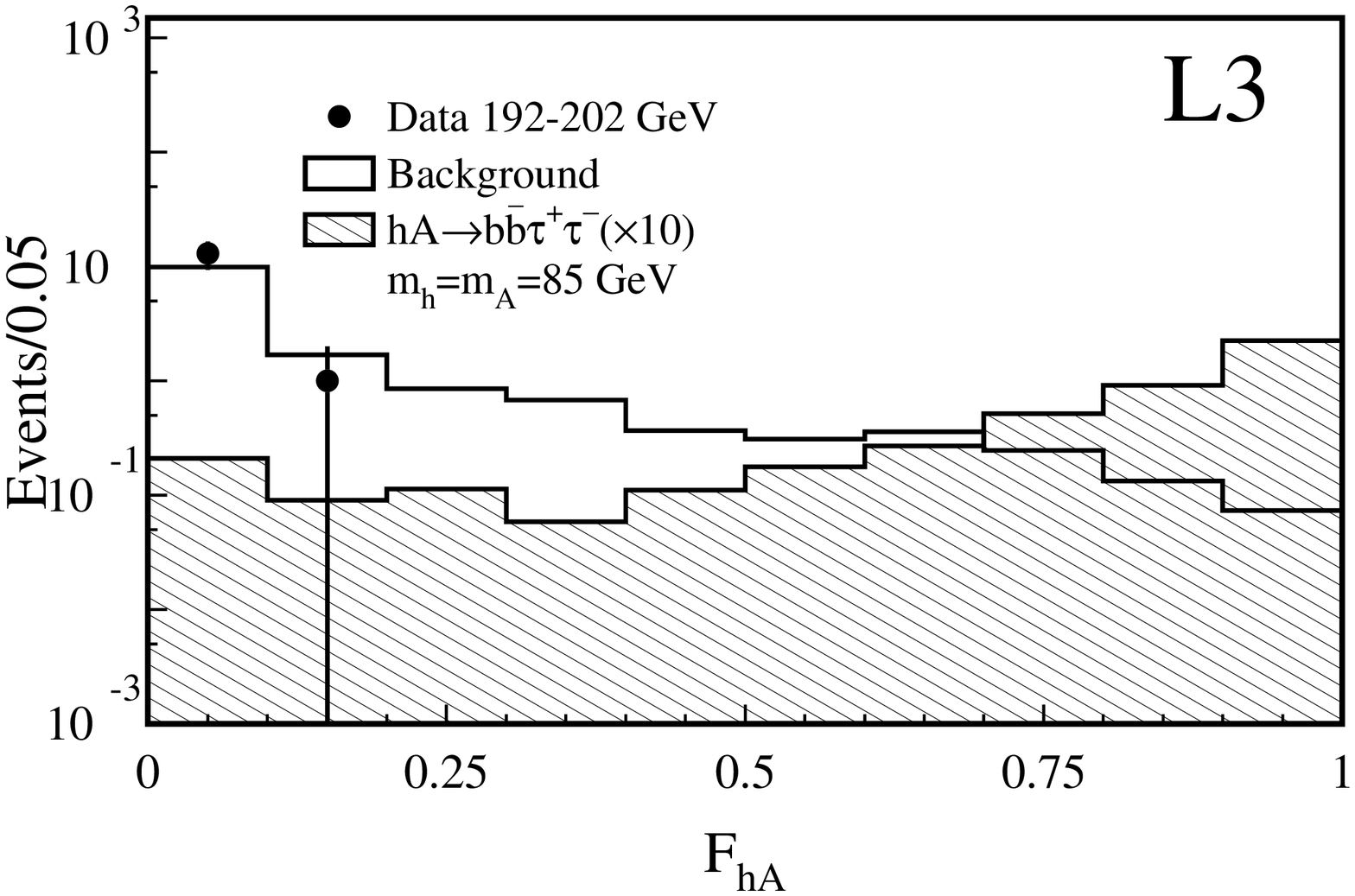}    
   \caption{Distribution of the discriminating variable of the
      $\hAtobbtt$ selection 
      in the hypothesis $\mh=m_A=85\GeV$ at $\tanb=30$.
         The points are the data collected at $\sqrt{s}=192-202\GeV$, 
         the open histogram is
         the expected Standard Model background and the hatched histogram 
         is the expected signal 
         for the \hA search multiplied by a factor 10.}
    \label{fig:fv_bbtt}
  \end{center}
\end{figure}
\end{center}

Events retained by both the \hA and the \hZ selections, are classified 
according to highest value of the likelihood, as \hA or \hZ candidates. 
An example of the distribution of the discriminating variable for the \hA
search is shown in Figure~\ref{fig:fv_bbtt}.
Good agreement between the observed data 
and the expected background is found.
The mass distribution for the \hA search after an additional 
cut on the $\rm{B_{Tag}}$ is shown in
Figure~\ref{fig:mass}b.
    
\begin{table}[htb]
  \begin{center}
    \begin{tabular}{|c|rr|rr|rr|rr|}
    \hline
 & \multicolumn{2}{|c|}{192\GeV} &\multicolumn{2}{|c|}{196\GeV}
 &\multicolumn{2}{|c|}{200\GeV} & \multicolumn{2}{|c|}{202\GeV} \\  
\hline
 & \hA  & \hZ  & \hA  & \hZ   & \hA  & \hZ   & \hA  & \hZ  \\
\hline\hline
Data                   &  2       & 2   & 6       & 7 & 5      & 7  &  3   &   3    \\
\hline
MC                  &  2.6     & 3.0 & 7.3     & 8.4 & 7.2    & 7.6 & 3.2    & 3.4  \\
\hline
\epemtoqq              & 0.3      & 0.5 & 0.8    & 1.4 & 0.9    & 0.8 & 0.4    & 0.3      \\
\epemtoWW              & 1.8      & 1.8 & 5.0     & 5.0 & 4.6    & 4.7 & 2.0    & 2.1 \\
\epemtoZZ              & 0.5      & 0.6 & 1.4     & 1.7  & 1.6    & 1.8  & 0.7     & 0.8  \\
$\mathrm{e^+e^-\rightarrow Ze^+e^-}$ & 0.0  & 0.1 & 0.1  & 0.2 & 0.1  &
 0.3 & 0.0  & 0.1 \\
\hline
$\varepsilon$(\hAtobbtt)   & 36\%   & 36\% & 36\%   & 36\% & 33\%   & 34\%  & 33\%   & 34\%   \\
$\varepsilon$(\hZtobbtt)   & 21\%   & 29\% & 21\%   & 29\%  & 17\%   & 29\% & 17\%   & 29\%  \\
$\varepsilon$(\hZtottqq)   & 20\% &  31\% & 20\% & 31\% & 20\% & 29\% & 20\% & 29\% \\
\hline
    \end{tabular}
    \caption{Number of events observed and expected in the tau
      selection. Efficiencies for the \hA signal
      are quoted for $\mA=\mh=85\GeV$. For the
      hZ signal, they are quoted for $\mh=100\;\GeV$ at 
      $\sqrt{\mathrm{s}}=192-196\;\GeV{}$ and for $\mh=105\;\GeV{}$
      at $\sqrt{\mathrm{s}}=200-202\;\GeV{}$.}
    \label{tab:bbtt_eff}
  \end{center}
\end{table}

%
%
\section{Results and Interpretation}
A good agreement between data and expected background, both in the
total number of events and in the shape of the distributions, is
observed in all the channels analysed.
The mass distributions of the events in the highest sensitivity region
are not compatible with a signal for any mass hypothesis. 
Therefore no evidence of the production 
of the \h and \A bosons is found. 

The results of the search for the \hA and the \hZ production are
interpreted in the framework of the constrained
MSSM (CMSSM) assuming unification of the scalar fermion masses, 
unification of the gaugino masses
and unification of the trilinear Higgs-fermion couplings at the GUT scale.
This choice has little impact on the phenomenology of the Higgs bosons 
but reduces
significantly the number of free parameters. The remaining free parameters
are \tanb, \mA, the gaugino mass
parameter, $M_2$, the universal scalar fermion mass, 
$m_0$, the common scalar quark trilinear
coupling, $A$, and the Higgs mixing parameter, $\mu$. 

Two benchmark scenarios ~\cite{carena_weiglein} are considered.
In the first one, termed ``maximal mixing'', 
the CMSSM parameters are chosen such that \mh acquires its maximal
value for any given value of \mA and \tanb.  
The second scenario corresponds to vanishing mixing 
in the scalar top sector and is referred to as 
``minimal mixing''.
  
The CMSSM parameters are chosen as follows:  
$m_0=1$\TeV, $\mu=-200$\GeV,
$M_2=200$\GeV.
The mass of the top quark is fixed to 175\GeV.  
The maximal mixing scenario is realised at  
$X_t=A-\mu\cot\beta=\sqrt{6}$\TeV, where $X_t$ is the parameter
which controls the mixing in the scalar top sector.
The minimal mixing corresponds to $X_t=0$.
Keeping these values fixed, a scan over the two remaining independent 
parameters, 
\tanb and \mA, is performed in each mixing scheme in the ranges:
0.5 $\le$ \tanb $\le$ 30 and 10\GeV $\le$ \mA $\le$ 1\TeV.

To set exclusion limits on the CMSSM parameters, the confidence level,
CL, that the expected signal is absent in the data, 
is calculated~\cite{ratio_method} for each point \tanbma of the scan. 
The full distributions of the discriminating variables, ${NN}$ and
${F}$, are used in this calculation.

Systematic and statistical uncertainties on the signal and background are 
evaluated using the same procedure as in the Standard Model Higgs search
~\cite{l3_sm_higgs_00_paper}.
The main sources of systematic uncertainties are detector resolution, 
selection procedures, theoretical
uncertainties and Monte Carlo statistics. 
The overall systematic uncertainties are estimated to be 4\% on
the predictions for the expected signal
and 10\% for the background events.
Bins of the final variables with a
signal-over-background ratio in the Monte Carlo of less than 
0.05 are not considered in the calculation of the CL.  
This cut is chosen to minimise the effect of
systematic uncertainties on the average CL as calculated from a large set of
Monte Carlo experiments.

\begin{figure}[htb]
  \begin{center}
    \includegraphics*[width=0.9\textwidth,bb=16 184 624 1038]{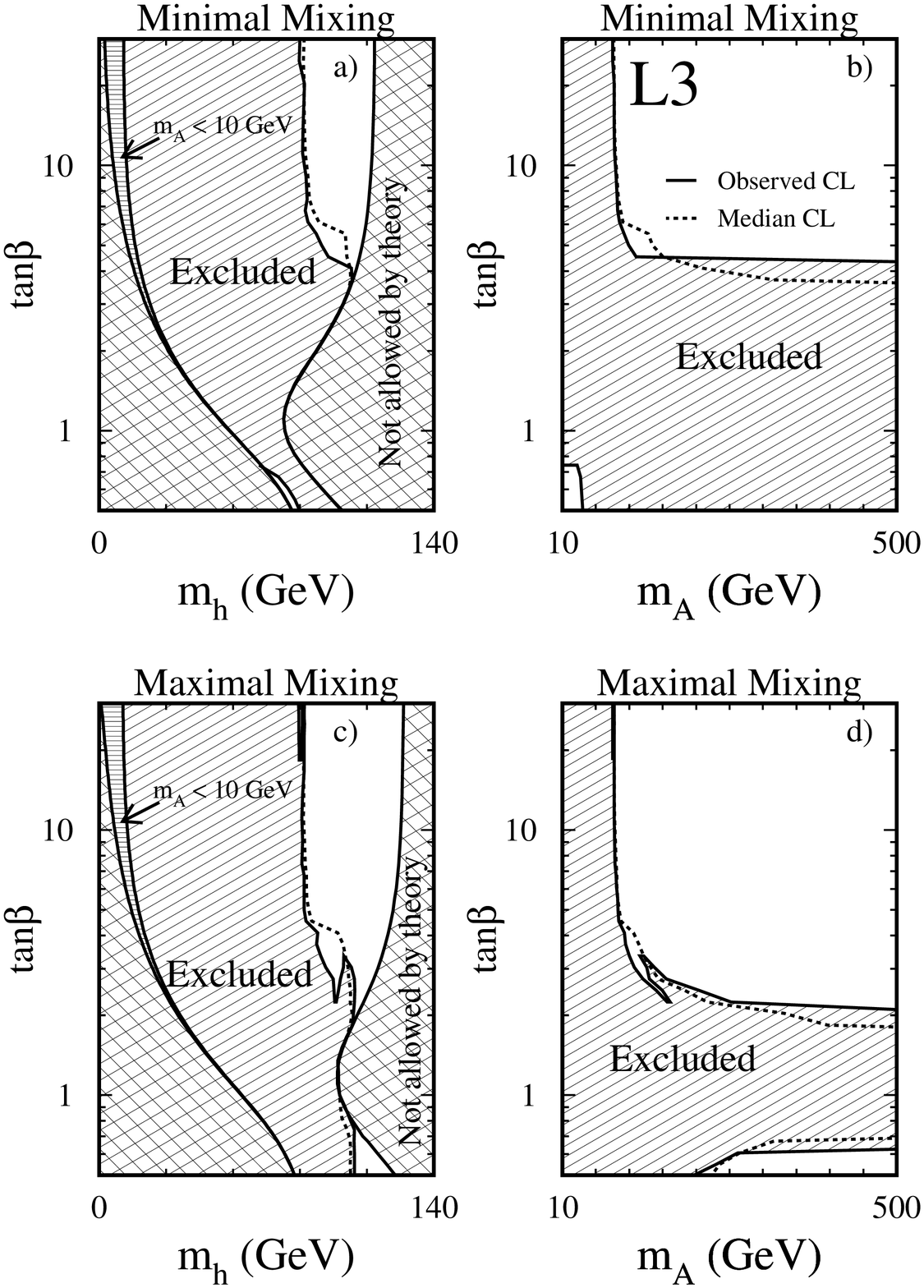} 
    \caption{Exclusion plots in the (\tanb,\mh) and (\tanb,\mA) planes at the
      95\% CL for the minimal and maximal mixing scenarios.  
      The  hatched area represents the exclusion and  
      the  crossed area is not
      allowed by the theory. The horizontal hatched area corresponds to 
      $\mA<10\;\GeV{}$ and was previously excluded at
      LEP \protect\cite{opal_2}.
      }
    \label{fig:limit}
  \end{center}
\end{figure}

The results of the MSSM Higgs search at lower $\sqrt{s}$~\cite{l3_1999_16}
are combined with those presented in
this paper.
Figure~\ref{fig:limit} shows the region of the (\tanb,\mh) 
plane and (\tanb,\mA) plane excluded by L3 for the maximal mixing and 
minimal mixing scenarios. 

For the CMSSM parameters considered and assuming \tanb greater than
0.8, this results in lower mass limits at the 95\% CL of:
\begin{displaymath}
  \mh > 83.4 \GeV, \;\;
  \mA > 83.8 \GeV,
\end{displaymath}
which compare to the  median expected 
limits in the absence of a signal of
$\mh > 85.6 \GeV{}$ and \mbox{$\mA > 85.7 \GeV{}$}.

 The exclusion plots for the minimal mixing scenario 
 present a small unexcluded area in the low \tanb region at low values
 of \mA where the decay $\mathrm{h\rightarrow AA}$
 is allowed but is not investigated among the signatures
 described above. 

For $0.8<\tanb<1.8$ values of \mA up to 
 1 TeV are ruled out for any mixing scenario
 allowing to exclude this \tanb region in the CMSSM, for the top mass 
$\lesssim175\GeV$.

%
%
\section{Acknowledgements}
We acknowledge the efforts of the engineers and technicians who
have participated in the construction and maintenance of L3 and
express our gratitude to the CERN accelerator divisions for the superb
performance of LEP.

%
%

%
%
\newpage
\typeout{   }     
\typeout{Using author list for paper 229 -- ? }
\typeout{$Modified: Nov 18 2000 by smele $}
\typeout{!!!!  This should only be used with document option a4p!!!!}
\typeout{   }
%
%
%
%
%
%

\newcount\tutecount  \tutecount=0
\def\tutenum#1{\global\advance\tutecount by 1 \xdef#1{\the\tutecount}}
\def\tute#1{$^{#1}$}
\tutenum\aachen            
\tutenum\nikhef            
\tutenum\mich              
\tutenum\lapp              
\tutenum\basel             
\tutenum\lsu               
\tutenum\beijing           
\tutenum\berlin            
\tutenum\bologna           
\tutenum\tata              
\tutenum\ne                
\tutenum\bucharest         
\tutenum\budapest          
\tutenum\mit               
\tutenum\debrecen          
\tutenum\florence          
\tutenum\cern              
\tutenum\wl                
\tutenum\geneva            
\tutenum\hefei             
\tutenum\seft              
\tutenum\lausanne          
\tutenum\lecce             
\tutenum\lyon              
\tutenum\madrid            
\tutenum\milan             
\tutenum\moscow            
\tutenum\naples            
\tutenum\cyprus            
\tutenum\nymegen           
\tutenum\caltech           
\tutenum\perugia           
\tutenum\cmu               
\tutenum\prince            
\tutenum\rome              
\tutenum\peters            
\tutenum\potenza           
\tutenum\riverside         
\tutenum\salerno           
\tutenum\ucsd              
\tutenum\sofia             
\tutenum\korea             
\tutenum\alabama           
\tutenum\utrecht           
\tutenum\purdue            
\tutenum\psinst            
\tutenum\zeuthen           
\tutenum\eth               
\tutenum\hamburg           
\tutenum\taiwan            
\tutenum\tsinghua          

{
\parskip=0pt
\noindent
{\bf The L3 Collaboration:}
\ifx\selectfont\undefined
 \baselineskip=10.8pt
 \baselineskip\baselinestretch\baselineskip
 \normalbaselineskip\baselineskip
 \ixpt
\else
 \fontsize{9}{10.8pt}\selectfont
\fi
\medskip
\tolerance=10000
\hbadness=5000
\raggedright
\hsize=162truemm\hoffset=0mm
\def\r{\rlap,}
\noindent

M.Acciarri\r\tute\milan\
P.Achard\r\tute\geneva\ 
O.Adriani\r\tute{\florence}\ 
M.Aguilar-Benitez\r\tute\madrid\ 
J.Alcaraz\r\tute\madrid\ 
G.Alemanni\r\tute\lausanne\
J.Allaby\r\tute\cern\
A.Aloisio\r\tute\naples\ 
M.G.Alviggi\r\tute\naples\
G.Ambrosi\r\tute\geneva\
% ...
\newpage
\begin{list}{A}{\itemsep=0pt plus 0pt minus 0pt\parsep=0pt plus 0pt minus 0pt
                \topsep=0pt plus 0pt minus 0pt}
\item[\aachen]
 I. Physikalisches Institut, RWTH, D-52056 Aachen, FRG$^{\S}$\\
 III. Physikalisches Institut, RWTH, D-52056 Aachen, FRG$^{\S}$
\item[\nikhef] National Institute for High Energy Physics, NIKHEF, 
     and University of Amsterdam, NL-1009 DB Amsterdam, The Netherlands
\item[\mich] University of Michigan, Ann Arbor, MI 48109, USA
\item[\lapp] Laboratoire d'Annecy-le-Vieux de Physique des Particules, 
     LAPP,IN2P3-CNRS, BP 110, F-74941 Annecy-le-Vieux CEDEX, France
\item[\basel] Institute of Physics, University of Basel, CH-4056 Basel,
     Switzerland
\item[\lsu] Louisiana State University, Baton Rouge, LA 70803, USA
\item[\beijing] Institute of High Energy Physics, IHEP, 
  100039 Beijing, China$^{\triangle}$ 
\item[\berlin] Humboldt University, D-10099 Berlin, FRG$^{\S}$
\item[\bologna] University of Bologna and INFN-Sezione di Bologna, 
     I-40126 Bologna, Italy
\item[\tata] Tata Institute of Fundamental Research, Bombay 400 005, India
\item[\ne] Northeastern University, Boston, MA 02115, USA
\item[\bucharest] Institute of Atomic Physics and University of Bucharest,
     R-76900 Bucharest, Romania
\item[\budapest] Central Research Institute for Physics of the 
     Hungarian Academy of Sciences, H-1525 Budapest 114, Hungary$^{\ddag}$
\item[\mit] Massachusetts Institute of Technology, Cambridge, MA 02139, USA
\item[\debrecen] KLTE-ATOMKI, H-4010 Debrecen, Hungary$^\P$
\item[\florence] INFN Sezione di Firenze and University of Florence, 
     I-50125 Florence, Italy
\item[\cern] European Laboratory for Particle Physics, CERN, 
     CH-1211 Geneva 23, Switzerland
\item[\wl] World Laboratory, FBLJA  Project, CH-1211 Geneva 23, Switzerland
\item[\geneva] University of Geneva, CH-1211 Geneva 4, Switzerland
\item[\hefei] Chinese University of Science and Technology, USTC,
      Hefei, Anhui 230 029, China$^{\triangle}$
\item[\lausanne] University of Lausanne, CH-1015 Lausanne, Switzerland
\item[\lecce] INFN-Sezione di Lecce and Universit\`a Degli Studi di Lecce,
     I-73100 Lecce, Italy
\item[\lyon] Institut de Physique Nucl\'eaire de Lyon, 
     IN2P3-CNRS,Universit\'e Claude Bernard, 
     F-69622 Villeurbanne, France
\item[\madrid] Centro de Investigaciones Energ{\'e}ticas, 
     Medioambientales y Tecnolog{\'\i}cas, CIEMAT, E-28040 Madrid,
     Spain${\flat}$ 
\item[\milan] INFN-Sezione di Milano, I-20133 Milan, Italy
\item[\moscow] Institute of Theoretical and Experimental Physics, ITEP, 
     Moscow, Russia
\item[\naples] INFN-Sezione di Napoli and University of Naples, 
     I-80125 Naples, Italy
\item[\cyprus] Department of Natural Sciences, University of Cyprus,
     Nicosia, Cyprus
\item[\nymegen] University of Nijmegen and NIKHEF, 
     NL-6525 ED Nijmegen, The Netherlands
\item[\caltech] California Institute of Technology, Pasadena, CA 91125, USA
\item[\perugia] INFN-Sezione di Perugia and Universit\`a Degli 
     Studi di Perugia, I-06100 Perugia, Italy   
\item[\cmu] Carnegie Mellon University, Pittsburgh, PA 15213, USA
\item[\prince] Princeton University, Princeton, NJ 08544, USA
\item[\rome] INFN-Sezione di Roma and University of Rome, ``La Sapienza",
     I-00185 Rome, Italy
\item[\peters] Nuclear Physics Institute, St. Petersburg, Russia
\item[\potenza] INFN-Sezione di Napoli and University of Potenza, 
     I-85100 Potenza, Italy
\item[\riverside] University of Californa, Riverside, CA 92521, USA
\item[\salerno] University and INFN, Salerno, I-84100 Salerno, Italy
\item[\ucsd] University of California, San Diego, CA 92093, USA
\item[\sofia] Bulgarian Academy of Sciences, Central Lab.~of 
     Mechatronics and Instrumentation, BU-1113 Sofia, Bulgaria
\item[\korea]  Laboratory of High Energy Physics, 
     Kyungpook National University, 702-701 Taegu, Republic of Korea
\item[\alabama] University of Alabama, Tuscaloosa, AL 35486, USA
\item[\utrecht] Utrecht University and NIKHEF, NL-3584 CB Utrecht, 
     The Netherlands
\item[\purdue] Purdue University, West Lafayette, IN 47907, USA
\item[\psinst] Paul Scherrer Institut, PSI, CH-5232 Villigen, Switzerland
\item[\zeuthen] DESY, D-15738 Zeuthen, 
     FRG
\item[\eth] Eidgen\"ossische Technische Hochschule, ETH Z\"urich,
     CH-8093 Z\"urich, Switzerland
\item[\hamburg] University of Hamburg, D-22761 Hamburg, FRG
\item[\taiwan] National Central University, Chung-Li, Taiwan, China
\item[\tsinghua] Department of Physics, National Tsing Hua University,
      Taiwan, China
\item[\S]  Supported by the German Bundesministerium 
        f\"ur Bildung, Wissenschaft, Forschung und Technologie
\item[\ddag] Supported by the Hungarian OTKA fund under contract
numbers T019181, F023259 and T024011.
\item[\P] Also supported by the Hungarian OTKA fund under contract
  numbers T22238 and T026178.
\item[$\flat$] Supported also by the Comisi\'on Interministerial de Ciencia y 
        Tecnolog{\'\i}a.
\item[$\sharp$] Also supported by CONICET and Universidad Nacional de La Plata,
        CC 67, 1900 La Plata, Argentina.
\item[$\diamondsuit$] Also supported by Panjab University, Chandigarh-160014, 
        India.
\item[$\triangle$] Supported by the National Natural Science
  Foundation of China.
\end{list}
}
\vfill


\newpage

\end{document}